\documentclass[reprint,aps,prl,superscriptaddress,showpacs,footinbib,longbibliography,amsmath]{revtex4-1}

\usepackage{graphicx}
\usepackage{dcolumn}
\usepackage{bm}

\usepackage{soul}
\setstcolor{red}

\usepackage{color}

\begin{document}
\title{Spin-liquid signatures in the quantum critical regime of pressurized CePdAl}
\author{M. Majumder}
\affiliation{Department of Physics, Shiv Nadar University, Gautam Buddha Nagar, Uttar Pradesh 201314, India}
\affiliation{Experimental Physics VI, Center for Electronic Correlations and Magnetism, University of Augsburg, 86159 Augsburg, Germany}

\author{R. Gupta}
\author{H. Luetkens}
\author{R. Khasanov}
\affiliation{Laboratory for Muon Spin Spectroscopy, Paul Scherrer Institut, 5232 Villigen PSI, Switzerland}

\author{O. Stockert}
\affiliation{Max Planck Institute for Chemical Physics of Solids, 01187 Dresden, Germany}

\author{P. Gegenwart}
\author{V. Fritsch}
\affiliation{Experimental Physics VI, Center for Electronic Correlations and Magnetism, University of Augsburg, 86159 Augsburg, Germany}

\date{\today}

\begin{abstract}

CePdAl is a prototypical frustrated Kondo lattice with partial long-range order (LRO) at $T_\mathrm{N}=2.7$~K. Previous bulk experiments under hydrostatic pressure found signatures for a quantum critical regime that extends from $p_\mathrm{c} \approx 0.9$~GPa, where LRO disappears, up to $\sim 1.7$~GPa. We employed extensive muon spin relaxation and rotation ($\mu$SR) experiments under pressure. The continuous and complete suppression of LRO at $p_\mathrm{c}$ is confirmed. Above $T_\mathrm{N}(p)$ and beyond $p_\mathrm{c}$, an additional crossover scale $T^\ast(p)$ characterizes the change from pure to stretched exponential relaxation in zero field. Remarkably $T^\ast(p)$ agrees with previously determined signatures of entropy accumulation above LRO. This coincidence microscopically evidences fluctuating frustrated spins at $T\leq T^\ast$ with spin-liquid behavior. Power-law divergences of the temperature and longitudinal field dependences of the relaxation rate, with time-field scaling, at pressures between $p_c$ and 1.7~GPa characterize this regime as quantum critical. 

\end{abstract}

\maketitle

\textcolor{blue}{\textit{Introduction.}} Due to the presence of competing interactions, strongly correlated electron systems react inherently susceptible to chemical composition or the application of pressure. The tuning of such non-thermal parameters often induces a quantum critical point (QCP), where magnetic order is continuously suppressed to zero temperature. Experimentally, non-Fermi liquid behavior was detected in the last decades for numerous quantum critical materials, most notably in the class of heavy-fermion metals, where it arises from the competition between the Kondo effect and the indirect exchange interaction between the $4f$-electrons ~\cite{Hilbert2007}. Its theoretical description is particularly challenging if a divergence of the quasiparticle mass indicates a severe failure of Fermi-liquid theory, beyond the framework of spin-density-wave criticality~\cite{Gegenwart2008}. Advanced scenarios like Fermi liquid fractionalization~\cite{Senthil2003}, breakdown of Kondo entanglement~\cite{Si2001} or strong-coupling critical Fermi liquid~\cite{Wolfle2017} were proposed. Within a "global phase diagram" unconventional quantum criticality as well as a metallic spin-liquid state with localized 4f moments was proposed to arise from strong quantum fluctuations, induced by low-dimensional and frustrated magnetic couplings ~\cite{Si2006,Vojta2008,Coleman2010}. Experimental investigation of quantum criticality in frustrated Kondo lattices is therefore of particular interest.

The hexagonal ZrNiAl structure features equilateral triangles of Zr atoms, forming a distorted Kagome lattice, offering the possibility of frustration when 4f moments are placed on the Zr lattice sites. For example, in YbAgGe, as function of applied magnetic field, a series of almost degenerate magnetically ordered states arises below 1~K with a quantum bicritical metamagnetic endpoint and pronounced non-Fermi liquid behavior in its vicinity~\cite{Dong2013,Tokiwa2013}. While in YbAgGe the 4f electrons are well localized due to a very weak Kondo interaction, isostructural CeRhSn and CeIrSn are Kondo lattices within the intermediate valence regime~\cite{Bando2000,Shimura2021}. For CeRhSn zero-field quantum criticality was detected, which is tunable by in-plane uniaxial pressure and related to geometrical frustration~\cite{Tokiwa2015,Kuchler2017}. CeIrSn features even stronger valence fluctuations, as recently confirmed spectroscopically~\cite{Shimura2021}. Despite a huge $T_\mathrm{K}\approx 480$~K, emergent antiferromagnetic (AF) correlations where detected below 1~K, suggesting that strong frustration counteracts Kondo singlet formation even in the intermediate valence regime~\cite{Shimura2021}.

We focus on isostructural CePdAl with $T_\mathrm{K}\approx 5$~K~\cite{Goto2002} which stands out of the class of geometrically frustrated Kondo systems due to its unique coexistence of AF order below $T_\mathrm{N}=2.7$~K~\cite{Kitazawa1994} with magnetic frustration, evidenced by the fact that one of three Ce moments in the crystallographic unit cell remains fluctuating down to 30~mK~\cite{Donni1996,Oyamada2008}. Powder neutron diffraction was described by a simplified model. Within the basal plane, ferromagnetic (FM) nearest neighbor interactions compete with next-nearest neighbor AF coupling, leaving Ising spins (along the c-axis) that form FM chains of Ce(1) and Ce(3) atoms, separated by fluctuating Ce(2) moments~\cite{Donni1996}. A detailed magnetic and thermodynamic investigation on single crystalline CePdAl under c-axis magnetic field found a scale $T_S(B)$, where magnetization $M(T)$ and entropy $S(B)$ display maxima~\cite{Lucas2017}. In the absence of frustration, such entropy accumulation should occur at the AF phase boundary, while for CePdAl it arises significantly above $T_\mathrm{N}(B)$, indicating a regime with spin-liquid behavior at $T_\mathrm{N}\leq T\leq T_S$~\cite{Lucas2017}. The AF order in CePdAl can be continuously suppressed by chemical substitution in the alloying series CePd$_{1-x}$Ni$_x$Al~\cite{Fritsch2014,Sakai2016,Huesges2017}. Thermodynamic evidence for quantum criticality was found and an additional entropy signature beyond the QCP arises, likely related to persistent correlations of the frustrated Ce(2) moments~\cite{Sakai2016}. The AF order can also be suppressed by hydrostatic pressure~\cite{Goto2002,Prokes2007,Zhao2019}, yielding a QCP at $p_c \approx 0.9$~GPa. Maxima in the temperature dependence of the ac-susceptibility were used to determine $T_S(p)$; interestingly it persists beyond the critical pressure up to about 1.7~GPa and was interpreted as paramagnetic phase with 4f electrons that neither order nor form a Fermi liquid \cite{Zhao2019,Zhang2018,Wang2021}.

In this Letter, we report detailed low-temperature muon spin relaxation/rotation ($\mu$SR) measurements under hydrostatic pressure performed at the Muon source of the Paul Scherrer Institute in Switzerland which provide microscopic evidence for a quantum-critical regime with spin-liquid properties in CePdAl.
All relevant details on samples, spectrometers and data analysis can be found in the supplemental material~\cite{SM}.

\begin{figure}
{\centering {\includegraphics[width=\linewidth]{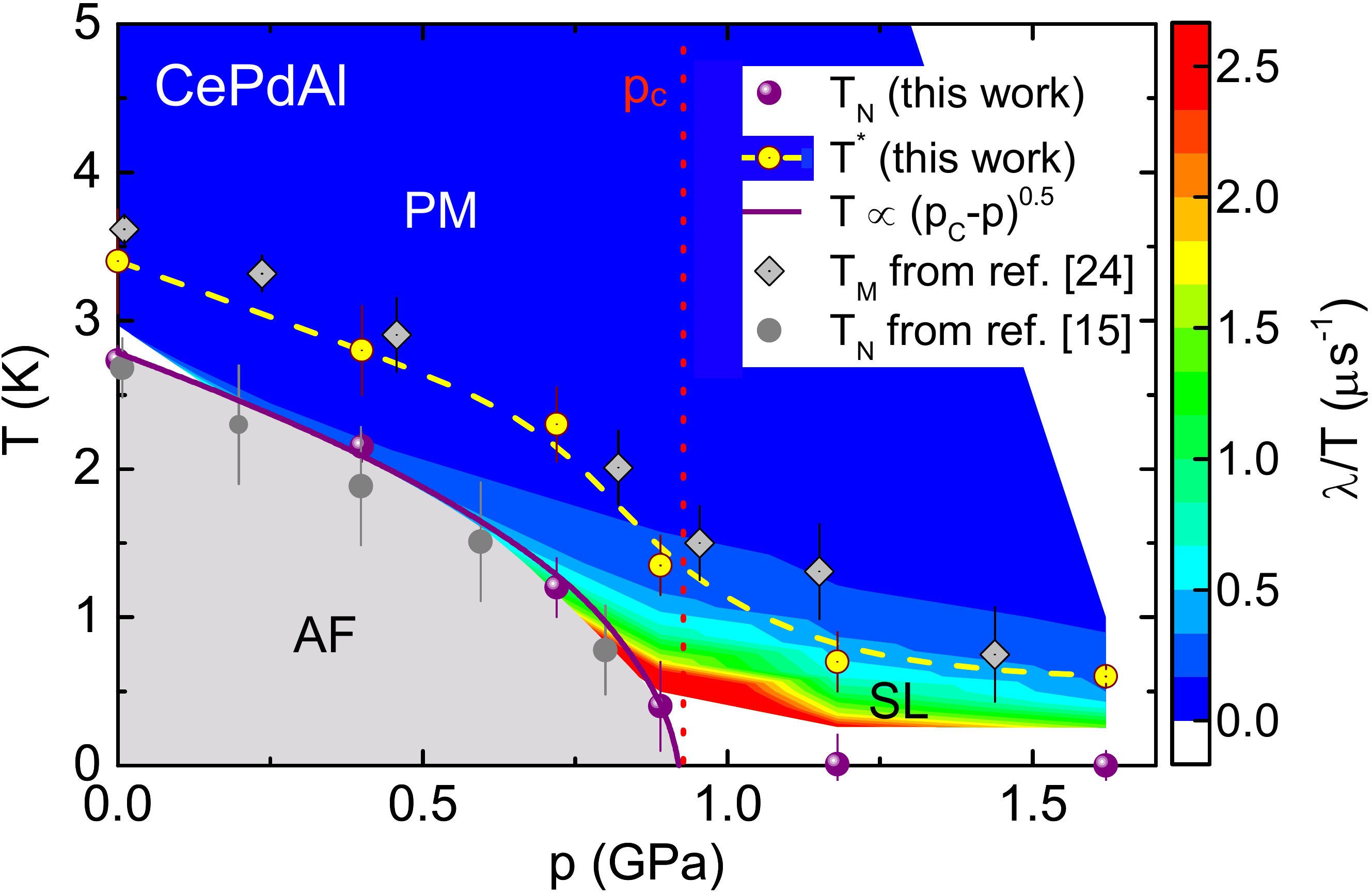}}\par} \caption{\label{fig:phase-diagram}Temperature-pressure magnetic phase diagram of CePdAl combined with a contour plot of the pressure- and temperature-dependent relaxation rate $\lambda(p,T)$ in CePdAl, plotted as $\lambda(p,T)/T$. Apart from the antiferromagnetically ordered (AF) and the paramagnetic (PM) phases there exists an extended region around and above pressure-induced quantum criticality (at $p_c$) with spin-liquid (SL) properties. The N\'eel temperatures $T_\mathrm{N}$ extracted from weak transverse-field $\mu$SR (purple symbols, error bars indicate the width of the transition) agree well with reported values (gray symbols) from susceptibility measurements \cite{Goto2002}. Open circles and open squares display $T^\ast(p)$ and the entropy accumulation line $T_S(p)$ (\cite{Zhao2019}, therein labeled as $T_\mathrm{M}$), respectively. The solid line corresponds to $(p_c-p)^{0.5}$ with $p_c = 0.92$\,GPa while the dashed line is a guide to the eye.} \label{structure}
\end{figure}

\begin{figure}
{\centering {\includegraphics[width=\linewidth]{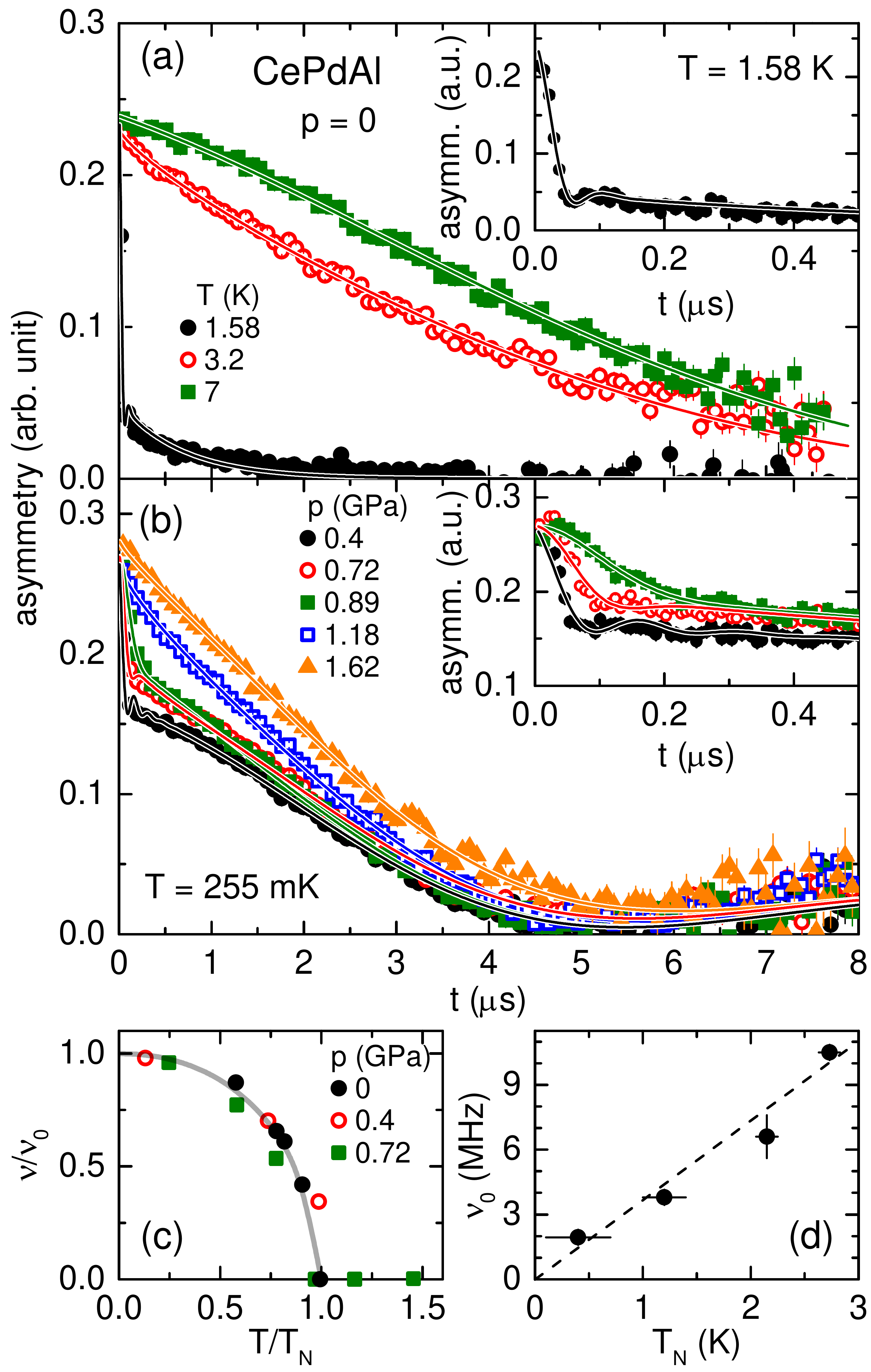}}\par} \caption{\label{fig:spectra} Zero-field (ZF) asymmetry spectra in CePdAl (a) for different temperatures at ambient pressure and (b) at 255~mK for different hydrostatic pressures. Insets in (a) and (b) zoom into the initial relaxation at early times. (c) Normalized muon oscillation frequency $\nu/\nu_0$ ($\nu_0 = \nu (T = 0)$) versus normalized temperature $T/T_\mathrm{N}$ for different pressures. (d) Muon oscillation frequency $\nu_0$ versus ordering temperature $T_\mathrm{N}$ indicating a linear dependence (dashed straight line). Solid lines denote fits to the data (for details see text).
} \label{structure}
\end{figure}

\textcolor{blue}{\textit{Nature of ordered state under pressure.}} 
Weak trans\-verse-field (wTF) measurements were performed to follow the N\'eel temperature $T_\mathrm{N}$ as function of pressure p~\cite{SM}. As displayed in Fig.\,\ref{fig:phase-diagram} $T_\mathrm{N}(p)$ agrees well with bulk measurements \cite{Goto2002} and is suppressed in a mean-field like behavior, i.e. $T_\mathrm{N}(p) \propto (p_c - p)^{0.5}$ with $p_c = 0.92$\,GPa.
Figs.~\ref{fig:spectra}(a) and (b) show selected zero-field (ZF) $\mu$SR spectra for ambient and various applied pressures. The ambient pressure spectrum at 1.58 K, i.e., below $T_\mathrm{N}$, contains a highly-damped oscillating component (cf. inset of Fig.~\ref{fig:spectra}(a)), that is absent in the paramagnetic state, in addition to the background relaxation. The best fit to the zero-field $\mu$SR spectra is obtained using:
\begin{eqnarray}
 & A(t) = A_0[f_\mathrm{sample}\lbrace\frac{2}{3}j_0(2\pi \nu t+\frac{\pi \phi}{180})e^{-\frac{(\sigma t)^2}{2}}+\frac{1}{3}e^{-\lambda_Lt}\rbrace \nonumber \\ 
 & + (1-f_\mathrm{sample})e^{-\lambda_\mathrm{Bkg} t}]
\end{eqnarray}
Here $A_0$ is the initial asymmetry and $\sigma$ the muon transverse depolarization rate (arising from a distribution of internal fields) forming a Gaussian envelope to the oscillating component with a frequency $\nu$ and phase $\phi$ (set to zero in our case). Furthermore, $\lambda_L$ is the longitudinal relaxation rate, $f_\mathrm{sample}$ the fraction of moments taking part in the LRO and $\lambda_\mathrm{Bkg}$ the relaxation rate of the background contribution. At 1.5 K, $f_\mathrm{sample} \approx 85\%$ which matches with the sample contribution found in wTF measurements. Thus, we kept this fraction fixed below $T_\mathrm{N}$. The need of the Bessel function ($j_0$) indicates an incommensurate long-range magnetic ordering consistent with neutron scattering and NMR experiments~\cite{Donni1996,Keller2002,Oyamada2008}. 

Data taken in the pressure cell were analyzed by the same function, taking into account the cell background~\cite{SM}. At the lowest temperature $T = 255$\,mK the oscillation frequency $\nu$ of the ZF asymmetry clearly decreases with pressure (cf. Fig.~\ref{fig:spectra}(b)) and beyond 0.89 GPa, no sign of long-range order (LRO) is found anymore. The normalized temperature dependence of the oscillation frequency for different pressures below $p_c$, displayed in Fig.\,\ref{fig:spectra}(c) and representing the order parameter, can be described by $\nu (T) = \nu_0[1-(T/T_\mathrm{N})^{\alpha'}]^{\beta'}$, following~\cite{Baker2011,Disseler2012},  with pressure independent exponents $\alpha' = 2.2$ and $\beta' = 0.5$ (solid line in Fig.\,\ref{fig:spectra}(c)). This suggest a persistent nature of the LRO up to the QCP, together with a continuous decrease of the ordered moment. The linear dependence of the ordered moment, being proportional to the muon oscillation frequency $\nu_0$, on $T_\mathrm{N}$ shown in Fig.~\ref{fig:spectra}(d)) is in line with earlier reports on CePd$_{1-x}$Ni$_x$Al and pressurized CePdAl~\cite{Huesges2017}.

\textcolor{blue}{\textit{Pressure dependence of spin dynamics in the para\-magnetic state.}} The ZF-$\mu$SR spectra in the paramagnetic state above $T_\mathrm{N}(p)$ are fitted by a static Gaussian Kubo-Toyabe function (for the nuclear moment contribution) multiplied by an exponential function with a stretched exponent $\beta$ for the dynamics of the electronic moments:
\begin{equation}
A(t) = A_0\left[\left(\frac{1}{3}+\frac{2}{3}\lbrace1-(\sigma_\mathrm{nucl} t)^2\rbrace e^{-\frac{(\sigma_\mathrm{nucl} t)^2 }{2}}\right) \times e^{-(\lambda t)^\beta}\right]
\end{equation}
\begin{figure}
{\centering {\includegraphics[width=0.9\linewidth]{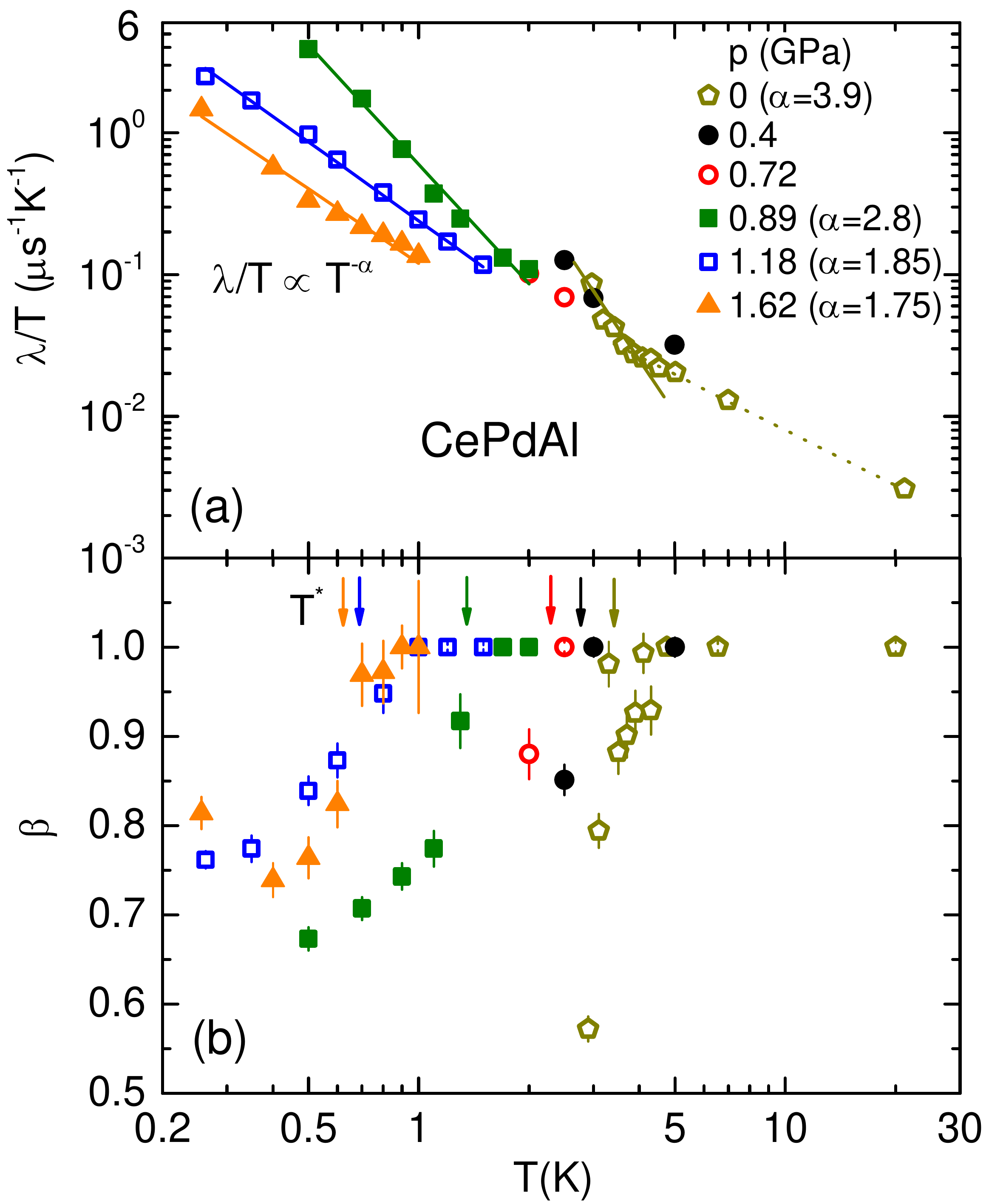}}\par} \caption{\label{fig:lambdaT}(a): Temperature dependence of the relaxation rate $\lambda$ in CePdAl plotted as $\lambda/T$ in zero field for various pressures. Solid lines represent power-law divergences ($T^{-\alpha}$). (b): Stretching exponent $\beta(T)$ at different pressures. Arrows indicate $T^\ast(p)$, where $\beta$ decreases below 1. For details see text.} \label{structure}
\end{figure}
We obtain a temperature independent $\sigma_\mathrm{nucl}\simeq 0.1325~\mu s^{-1}$, while the evolution of the relaxation rate $\lambda$ and stretching exponent $\beta$ with temperature and pressure is depicted in Fig.~\ref{fig:lambdaT}. We first discuss the evolution of $\beta$. At ambient pressure, a change from purely exponential ($\beta=1$) to stretched exponential ($\beta<1$) behavior is found below about 3.5~K. A transition to stretched exponential behavior is characteristic for disordered (fluctuating) moments and is regularly found in spin-glasses \cite{Campbell1994} and frustrated magnets \cite{Yaouanc2008,Li2016}. Note, that we do not see any indication of static fields from frozen moments at any pressure and temperature in CePdAl. In addition, the stretching exponent is far larger than $1/3$, characteristic for a spin-glass at the freezing temperature~\cite{Campbell1994}. We therefore associate the stretched exponential behavior at $T\leq T^\ast(p)$ (cf. the arrows in Fig.~\ref{fig:lambdaT}(b)) with the development of spin-liquid type correlations. In this temperature range quite strong, but very short-range AF correlations are visible in neutron scattering~\cite{Huesges2017}.
Most importantly, $T^\ast(p)$ agrees very well with the previously determined entropy accumulation $T_S(p)$ from bulk magnetic susceptibility. This provides microscopic evidence for frustrated magnetism over a wide parameter range, enclosing the LRO phase. Further information is contained in the temperature dependence of the relaxation rate $\lambda/T=1/T_1T \propto \chi''(q,\omega)$, where $\chi''(q,\omega)$ is the imaginary part of the dynamical spin susceptibility. The strong increase of $\lambda/T$ towards low $T$, shown for various pressures in Fig.~\ref{fig:lambdaT}(a), thus indicates the (critical) slowing down of the spin fluctuations. At and beyond $p_c$ the divergence holds to lowest measured $T$, indicative of quantum critical behavior, although we do not observe a universal, i.e., pressure-independent, critical exponent. As visualized in the contour plot (Fig.~\ref{fig:phase-diagram}), it is the regime beyond $p_c$, for which this divergence in $\lambda/T$ vs. $T$ is most prominent, providing microscopic evidence of a quantum-critical regime with spin-liquid behavior that extends up to at least $1.75 p_c$.

\begin{figure}
{\centering {\includegraphics[width=\linewidth]{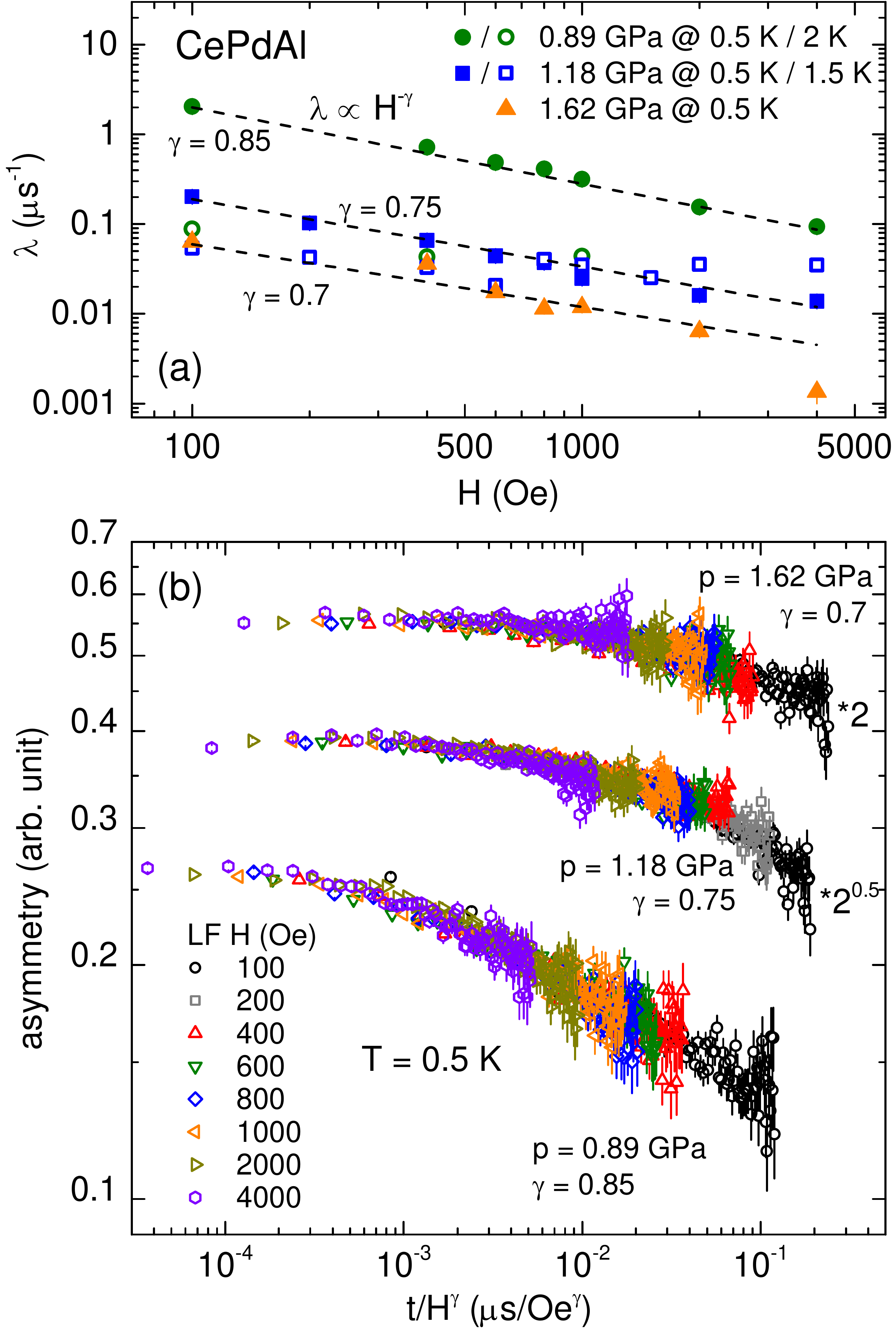}}} \caption{(a): Longitudinal field dependence of the $\mu$SR relaxation rate $\lambda$ in CePdAl at different pressures and temperatures (dashed lines are fits according to  $\lambda \propto H^{-\gamma}$ for $T = 0.5$\,K). (b) Time-over-field $t/H^{\gamma}$ scaling of the asymmetry spectra at $0.5$\,K for different pressures. The data for different pressures are shifted vertically by a factor $\sqrt{2}$ with respect to each other.} \label{fig:lambdaH}
\end{figure}

\textcolor{blue}{\textit{Quantum criticality close to $p_c$ and beyond.}} To characterize quantum critical fluctuations indicated by the divergent relaxation rate, we also performed longitudinal field (LF) $\mu$SR measurements. Fig.~\ref{fig:lambdaH}(a) shows the LF dependence of $\lambda$ at several fixed temperatures for three different pressures ranging from approx. $p_c$ to $1.75p_c$. In all cases, a power-law divergence towards zero-field is found, similar as recently reported for the nearly ferromagnetic metal YFe$_2$Al$_{10}$, where it was shown to resemble the dynamic structure factor $S(\omega,T)$ at the $\mu$SR frequencies~\cite{Huang2018}. With increasing temperature and pressure the exponent $\gamma$ in $\lambda \propto H^{-\gamma}$ is only weakly reduced (cf. Fig.~\ref{fig:lambdaH}(a)). Since $\mu$SR probes the Fourier transform of the spin-spin autocorrelation function $q(t)=\langle\textbf{S}_i(t).\textbf{S}_i(0)\rangle$, the nature of $q(t)$ (whether it follows a power-law or stretched exponential dependence) can be unambiguously identified by  time-field scaling of the $\mu$SR asymmetry. Fig.~\ref{fig:lambdaH}(b) shows $A(t/H_\mathrm{LF}^\gamma)$ scaling over three decades with consistent exponents $\gamma$ for the three pressures. Generally, $\gamma<1$ signifies a power-law behavior of $q(t)$, unlike $\gamma>1$ for a stretched exponential behavior in inhomogeneous systems. This strongly indicates that the spin dynamics is cooperative in the `spin-liquid' regime at $p_c \leq p<1.75 p_c$. Furthermore, the LF experiments also signify that the correlations are dynamic, since even at a field of about $2000$\,Oe no decoupling of the relaxation channels occurs.

\textcolor{blue}{\textit{Discussion.}} Generically, quantum criticality shrinks to a narrow control parameter regime at low temperatures. Quantum phase transitions can be smeared by disorder, resulting in a gradual evolution from LRO to a cluster glass with power-law divergences, as observed in
CePd$_{1-x}$Rh$_x$~\cite{Westerkamp2009}. Extended non-Fermi liquid regimes have also been observed when the transition changes its nature to first-order (discontinuous) near the critical point, cf. e.g. the cases in Sr$_{1-x}$Ca$_x$RuO$_3$ or MnSi under pressure~\cite{Uemura2007,Pfleiderer2007}. Clear absence of spin freezing, continuous suppression of the order parameter with pressure and divergence of the relaxation rate upon cooling exclude these scenarios for CePdAl. The onset of stretched exponential relaxation at $T<T^\ast(p)$ agrees with the entropy accumulation line $T_S(p)$, previously associated with spin-liquid correlations \cite{Zhao2019}. In this regime, the Ce-sites start to become inequivalent as evidenced by neutron diffraction~\cite{Huesges2017}. It is therefore obvious to ask, whether the stretched exponential behavior results from the superposition of $\mu$SR spectra from different muon stopping sites. However, below $T_\mathrm{N}(p)$, the muon stopping site(s) are obviously all sensitive to the static internal field from LRO, since the remaining paramagnetic signal can be associated with the background. This strongly suggests, that the stretched exponential relaxation indeed results from dynamical spin-liquid correlations and not from static spin disorder.

\textcolor{blue}{\textit{Conclusions.}} Detailed muon spin relaxation and rotation experiments under pressure and down to very low temperatures provide microscopic evidence for a quantum critical point in the hexagonal Kondo lattice \mbox{CePdAl} that is strongly affected by geometrical frustration. An extended regime with spin-liquid behavior is found,
with divergent relaxation rate and quantum-critical time over field scaling.

\textcolor{blue}{\textit{Acknowledgment.}} This work was supported by the German Research Foundation (DFG) through the research unit FOR 960 and via the Project No. 107745057 (TRR 80). V.F. acknowledges support by the Freistaat Bayern through the Programm f\"ur Chancen\-gleich\-heit f\"ur Frauen in Forschung und Lehre.

\textcolor{blue}{\textit{Author contributions.}} V.F. and O.S. initiated the project and prepared the sample. M.M. suggested and planned the $\mu$SR measurements. M.M., R.G., H.L. and R.K. performed the measurements. M.M., R.G., H.L., O.S., P.G. and V.F. discussed the results and analyzed the data. M.M., O.S. and P.G. wrote the manuscript. All authors revised and approved the manuscript.

\end{document}


\title{Supplemental material for "Spin-liquid signatures in the quantum critical regime of pressurized CePdAl"}
\author{M. Majumder}
\affiliation{Department of Physics, Shiv Nadar University, Gautam Buddha Nagar, Uttar Pradesh 201314, India}
\affiliation{Experimental Physics VI, Center for Electronic Correlations and Magnetism, University of Augsburg, 86159 Augsburg, Germany}

\author{R. Gupta}
\author{H. Luetkens}
\author{R. Khasanov}
\affiliation{Laboratory for Muon Spin Spectroscopy, Paul Scherrer Institut, 5232 Villigen PSI, Switzerland}

\author{O. Stockert}
\affiliation{Max Planck Institute for Chemical Physics of Solids, 01187 Dresden, Germany}

\author{P. Gegenwart}
\author{V. Fritsch}
\affiliation{Experimental Physics VI, Center for Electronic Correlations and Magnetism, University of Augsburg, 86159 Augsburg, Germany}

\date{\today}

\begin{abstract}

\end{abstract}

\maketitle

\section{Muon Spin Relaxation or Rotation Experiments}
Muon spin relaxation or rotation measurements at ambient pressure and under pressure have been performed on the GPS and GPD spectrometers of the Swiss Muon Source (S$\mu$S) at the Paul Scherrer Institute, Switzerland. The experiments have been carried out on approx. $2$\,g of CePdAl powder ground from single crystalline pieces. Other pieces of the Czochralski grown CePdAl single crystals have been thoroughly characterized \cite{Sakai2016,Lucas2017,Huesges2017}. 

A low-background double-walled CuBe/MP35N pressure cell has been used for the measurements under pressure. The details of this pressure cell can be found in refs. \cite{Rustem16,Rustem17}. Daphne7373 oil served as pressure transmitting medium to ensure hydrostatic conditions. Pressure was applied at room temperature and the resulting pressure at low temperatures was measured by monitoring the pressure-induced shift of the superconducting transition temperature of a little piece of indium also located in the pressure cell. The measured pressures have an error of about $0.05$\,GPa.

\begin{figure}[h]
{\centering {\includegraphics[width=0.9\linewidth]{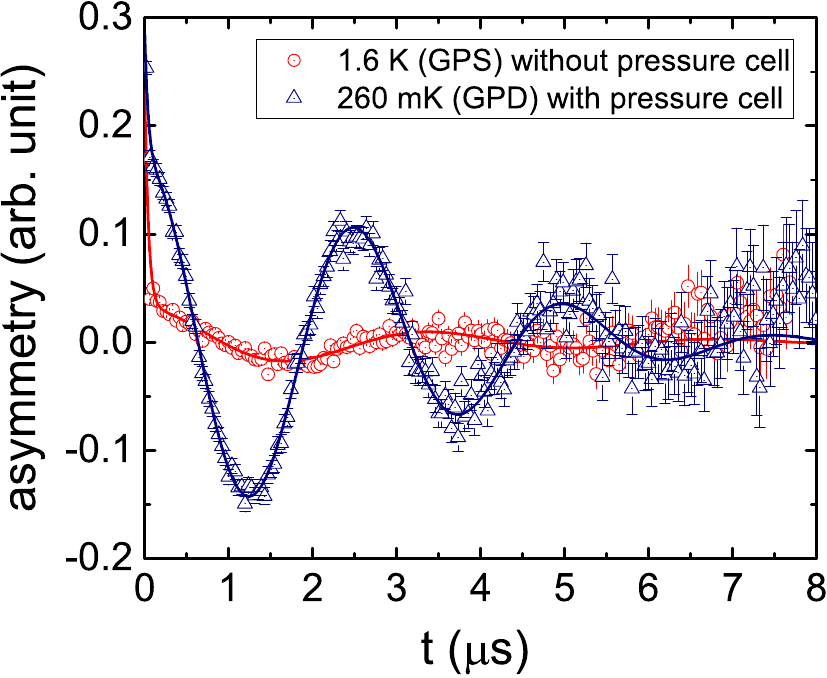}}\par} \caption{\label{fig:wTFraw} wTF $\mu$SR spectra ($\mu_0H = 20$\,G) in CePdAl at 1.6 K measures on GPS without the pressure cell and at 260 mK on GPD with the pressure cell. The line indicates the fit using equation (1).} \label{structure}
\end{figure}

\section{Weak transverse-field $\mu$SR measurements} 
All weak transverse-field (wTF) measurements on CePdAl have been performed in a field of $\mu_0H = 2$\,mT.
Fig.\,\ref{fig:wTFraw} displays some wTF data taken well below the ordering temperature and at ambient pressure on both, GPS (without pressure cell) and on GPD (with pressure cell). They clearly indicate  
some background contribution which does not get ordered. In the case of GPS the presence of such a background contribution may arise from the sample holder (which can contribute about 5-15\% of the total signal) and also from a small amount of impurities (e.g. CeO$_2$ on the surface of the sample) in CePdAl. Indeed, a contribution of up to 15\% resulting from background (muons stopping in sample holder or cryostat walls) has been found before for measurements on the GPS spectrometer \cite{Spehling13}. 

It should be noted that although in the partially frustrated antiferromagnetically ordered state of CePdAl $1/3$ of the Ce moments remain disordered down to lowest temperature, one would not expect $1/3$ of the muon stopping sites experiencing a zero local magnetic field. This is because the muon stopping sites are away from the Ce positions and there is no reason why the superposition of the dipolar fields should cancel at the muon stopping sites.

Hence, for our wTF measurements on GPS at ambient pressure the following equation has been used to describe the time-dependent spectra
\begin{equation}
A(t) = A_0 [(1-f) \cos (2 \pi \nu t+\dfrac{\pi \phi}{180})e^{-\lambda_{PM} t}  +fe^{-\lambda_{AF}t}]
\end{equation}
where $f$ indicates the signal fraction from antiferromagnetic order of CePdAl (with a damping rate $\lambda_{AF}$) and the rest, $1-f$, is the non-magnetic/paramagnetic contribution of sample and background (sample holder, cryostat walls) with $\lambda_{PM}$ as damping rate and $\nu$ as frequency. From fits to the data at lowest temperatures (solid line in Fig.\,S\ref{fig:wTFraw}), i.e. well below $T_\mathrm{N}$, the antiferromagnetic volume fraction $f$ to the total signal was found to be $f_0 = f(T\rightarrow 0) \approx 85$\% indicating a paramagnetic background contribution of $\approx 15$\% surviving down to the lowest temperature.

For measurements under applied hydrostatic pressure, a contribution from the (non-magnetic) pressure cell has been incorporated and the fits (cf. Fig.\, S\ref{fig:wTFraw}) yield approximately 50\% contribution from the pressure cell. Such a fraction is expected for the pressure cell used in our experiment \cite{Rustem16,Rustem17}. 
In addition to the pressure cell contribution, a 15\% fraction originating from non-magnetic/paramagnetic background was determined in our fits at the lowest pressure of $0.4$\,GPa and kept constant for all other pressures. This leaves around $f_0\approx 35$\% of the total signal as antiferromagnetic sample contribution well below $T_\mathrm{N}$ at $0.4$\,GPa.

\begin{figure}
{\centering {\includegraphics[width=\linewidth]{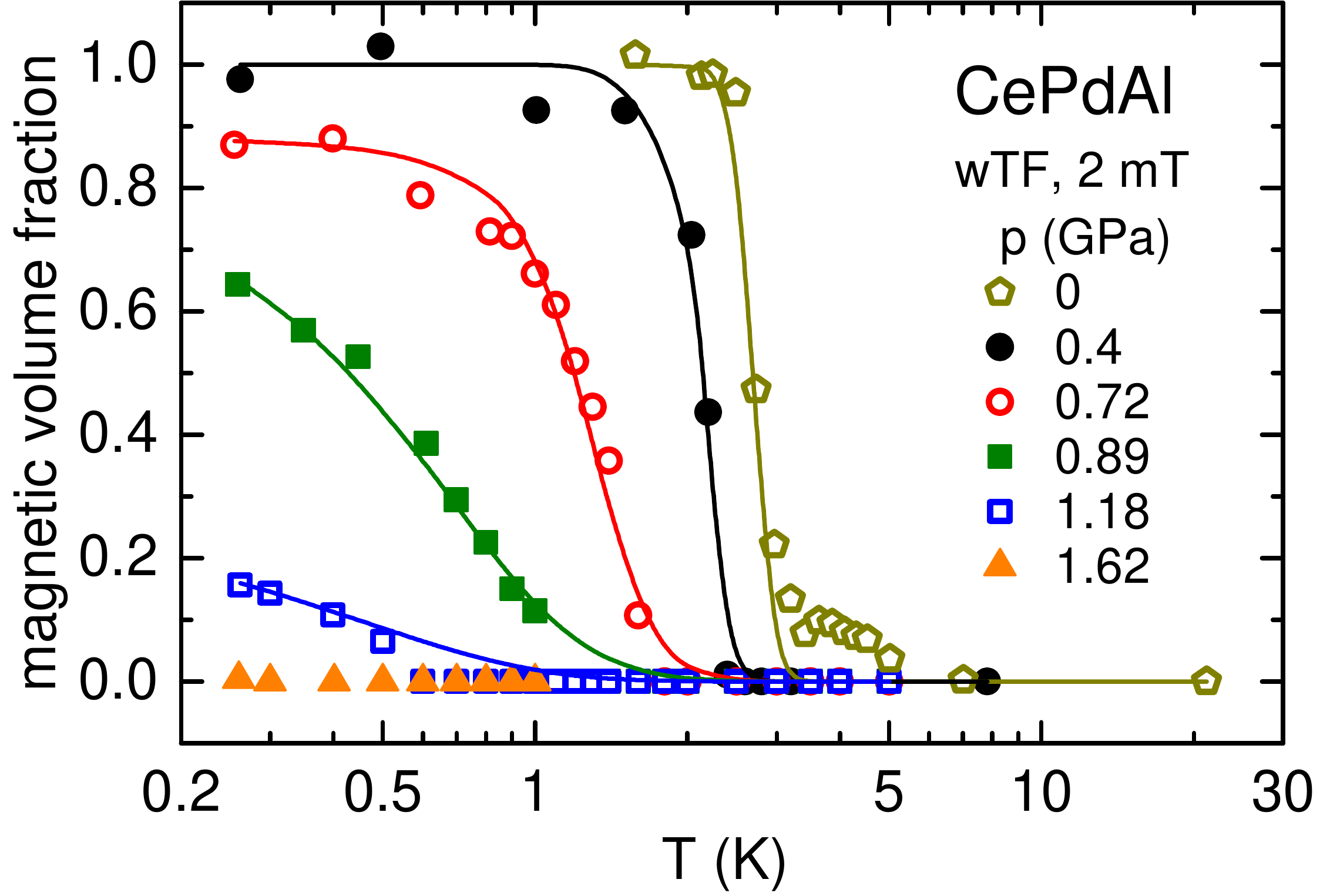}}\par} \caption{\label{fig:wTF} Temperature dependence of the normalized antiferromagnetically ordered volume fraction $f$ in CePdAl derived from wTF measurements at various pressures. Data are normalized to the respective $f_0$ for ambient conditions and under pressure (see text). The solid lines denote fits of Eq.\,\ref{sigmod} to the data.} \label{structure}
\end{figure}

The wTF measurements allow to extract the temperature dependence of the magnetic volume fraction. The antiferromagnetically ordered sample contribution to the total sample asymmetry, i.e. normalized to $f_0$, at various pressures is shown in Fig.~S\ref{fig:wTF} and was fitted by a simple sigmoidal function

\begin{equation}
f(T)/f_0 = \frac{2 f_0(T \rightarrow 0)}{1+\exp(\frac{T-T_\mathrm{N}}{\Delta T})}
\label{sigmod}
\end{equation}

to just estimate the ordering temperature $T_\mathrm{N}(p)$. Here, $f_0(T \rightarrow 0)$ is the maximum magnetic volume fraction of the sample at the respective pressure, $\Delta T$ is the width of the transition. The derived $T_\mathrm{N}(p)$ agree well with data from previous bulk experiments~\cite{Goto2002,Zhao2019} as can be seen in Fig.\,1 of the main text.

\begin{figure}
{\centering {\includegraphics[width=0.9\linewidth]{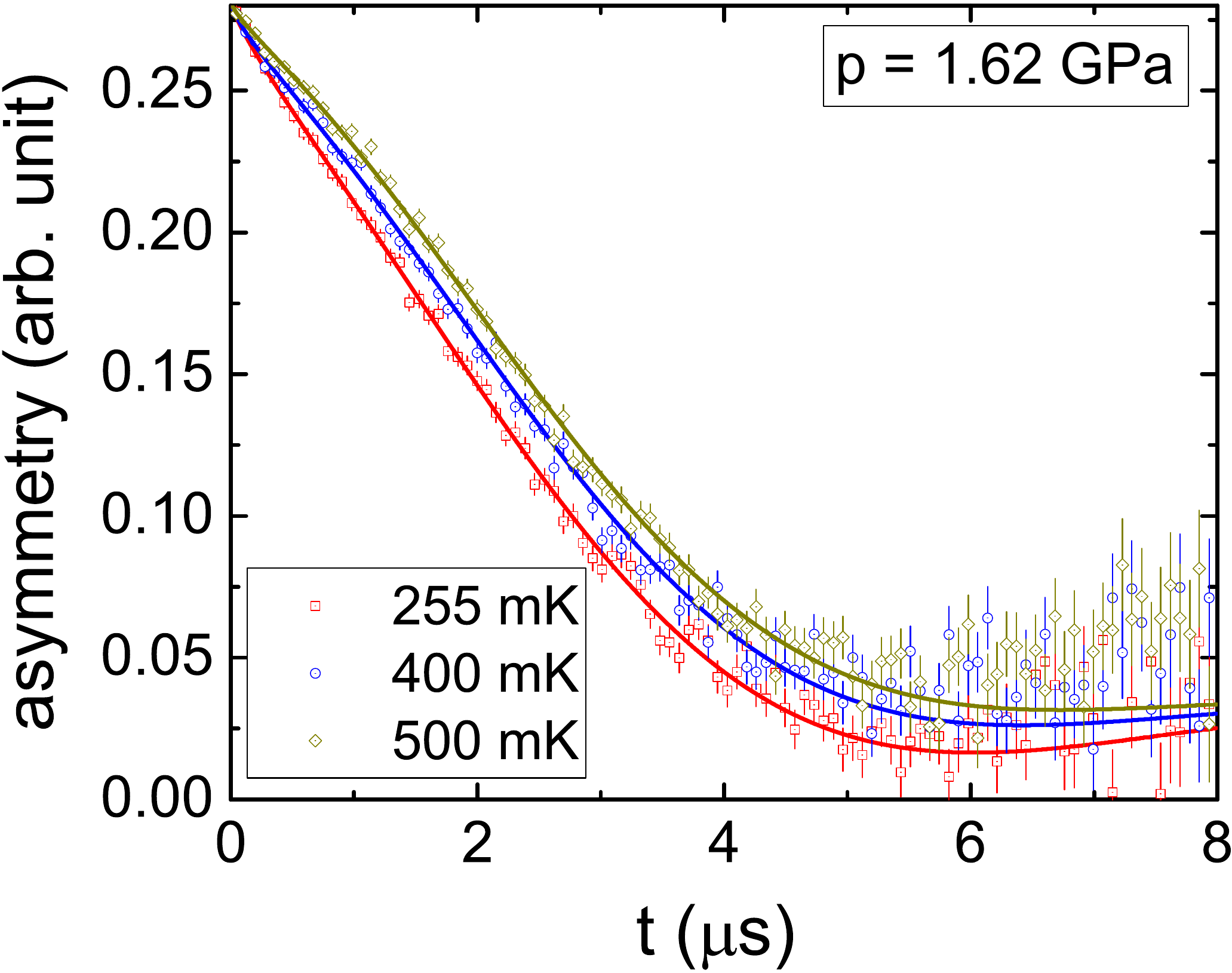}}\par} \caption{\label{fig:lambdaT} Temperature dependence of the ZF $\mu$SR spectra in CePdAl measured under a pressure of $p = 1.62$\,GPa. Solid lines represent fits to the data.} \label{structure}
\end{figure}

\section{Zero-field and longitudinal-field $\mu$SR measurements}
To fit the data of the zero-field (ZF) measurements under pressure we have used the following equation to include the pressure cell contribution

\begin{equation}
A_{total}(t) = F_{PC}A_{PC}(t)+F_SA(t)
\end{equation}  
where $F_{PC}$ and $F_S$ are the pressure cell and the sample contributions respectively. $A_{PC}$ corresponds to the function mentioned in equation (1) of reference \cite{Rustem17} for the pressure cell and $A(t)$ corresponds to equation 1 or 3 in the main manuscript to describe the antiferromagnetically ordered or paramagnetic states respectively. The individual values of the parameters corresponding to the pressure cell contribution are also depicted in Figure 8 of reference \cite{Rustem17}. As already mentioned above $F_{PC}$ was found to be 50\% as expected for the CuBe pressure cell \cite{Rustem16,Rustem17}.   
The raw ZF $\mu$SR spectra, shown in Fig.\,S3 for $p = 1.62$\,GPa at different temperatures, clearly indicates an increase of the relaxation rate $\lambda$ when lowering the temperature.

\begin{figure}
{\centering {\includegraphics[width=\linewidth]{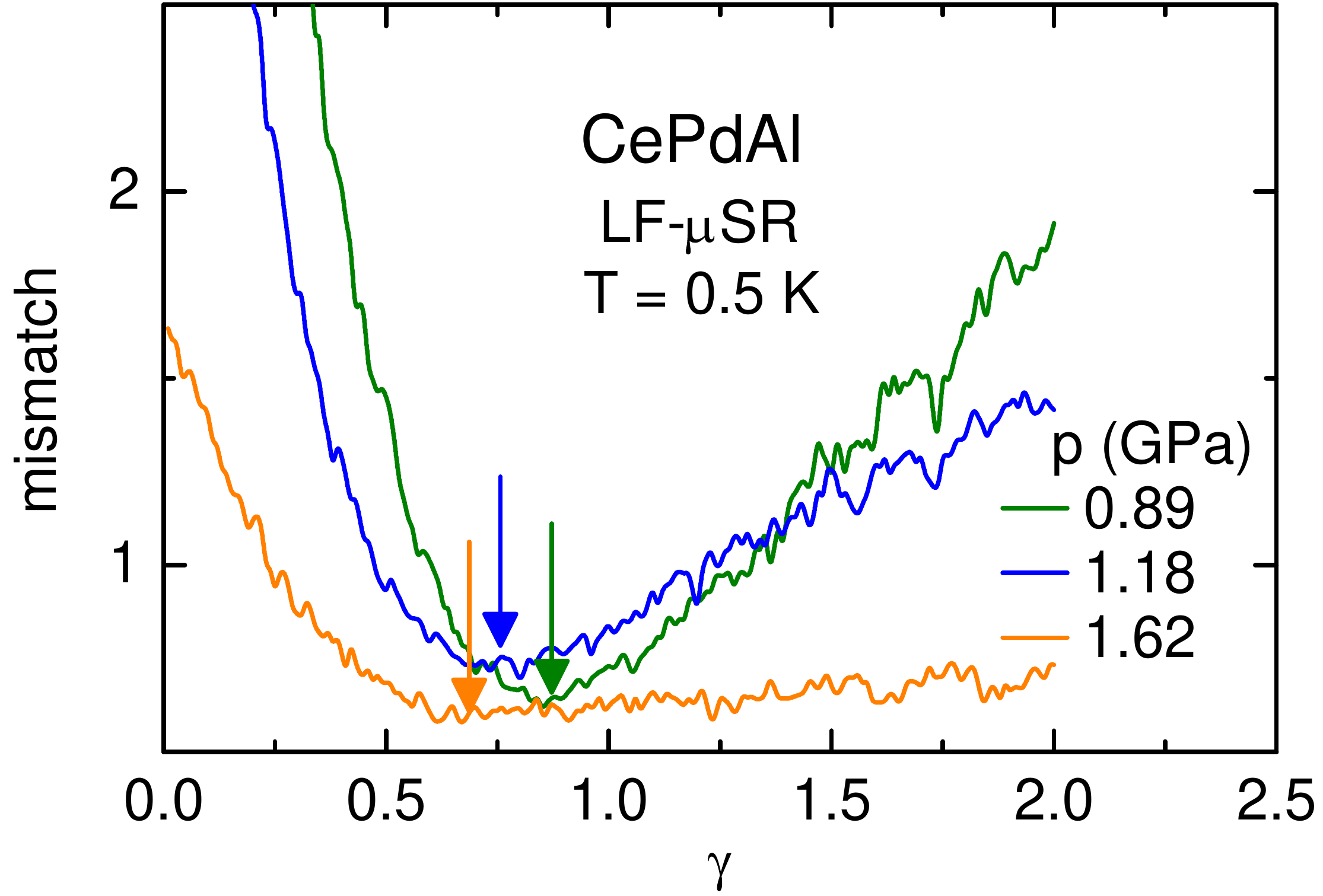}}\par} \caption{\label{fig:lambdaT} Mismatch functions for different pressures in CePdAl. The arrow indicates the lowest value of the exponent $\gamma$. For details see text.} \label{structure}
\end{figure}

Longitudinal-field (LF) experiments have been performed at different pressures ($p = 0.89$, $1.18$ and $1.62$\,GPa) at low temperatures ($T = 0.5$ and $2$\,K). The data points from all fields up to $t=6$\,$\mu$s (the data above $t=6$\,$\mu$s were not included due to their large error bars) were arranged with increasing $t/H^\gamma$ for every value of $\gamma$. An empirical data mismatch function was calculated by taking the difference between the neighboring points and weighing them by the corresponding error bars. The mismatch function can be defined as
\begin{equation}
M = \dfrac{1}{N}\sum_i^N \dfrac{(A_i-A_{i+1})^2}{(\delta_i-\delta_{i+1})^2}
\end{equation}
where $N$ is the number of data points, and $A_i$ and $\delta_i$ correspond to the asymmetry and error bar of the $i$-th data point, respectively. The lowest value of the mismatch function is obtained at $\gamma=0.85$, $0.75$ and $0.7$ for $p = 0.89$, $1.18$ and $1.62$\,GPa respectively. As seen in Fig.\,4(b) the universal scaling relation follows over at least three orders of magnitude in $t/H^{\gamma}$. The scaling relation has further been verified in the double logarithmic plot of $\lambda$ vs. $H$ shown in Fig.\,4(a) of the main manuscript indicating $\lambda \propto H^{-\gamma}$ at $T = 0.5$\,K with the same values for the exponent $\gamma$.